\documentclass[useASM]{mn2e}

\usepackage{graphicx}

\begin{document}

\title[Effect of magnetic field on NAE]
{Effect of magnetic field on neutrino annihilation efficiency in gamma-ray bursts}

\author[Du et al.]
{Shuang Du$^{1,4,5}$\thanks{E-mail: dushuang\_gxu@163.com}, Fang-Kun Peng$^{2,3,4,5}$, Guang-Bo Long$^{1}$, Miao Li $^{1}$\thanks{E-mail: limiao9@mail.sysu.edu.cn}\\
 $^1$Department of Physics and Astronomy, Sun Yat-Sen University, Zhuhai 519082, China\\
 $^2$Guizhou Provincial Key Laboratory of Radio Astronomy and Data Processing, Guizhou Normal University, Guiyang 550001, China\\
 $^3$Department of Physics And Electronics, Guizhou Normal University, Guiyang 550001, China\\
 $^4$GXU-NAOC Center for Astrophysics and Space Sciences, Department of Physics, Guangxi University, Nanning 530004, China\\
 $^5$Guangxi Key Laboratory for Relativistic Astrophysics, Nanning, Guangxi 530004, China}
 \maketitle

\label{firstpage}
\begin{abstract}
Neutrino annihilation process on a hyperaccreting disk is one of the leading models to explain the generation of relativistic jets of gamma-ray bursts (GRBs).
The neutrino annihilation efficiency (NAE) has been widely studied in black hol-accretion disc (BH¨Cdisc) system, and
the published results are mutually corroborated. 
However, there is still uncertainty regarding the NAE of neutron star-accretion disc (NS-disc) system 
because of complicated microphysics processes and effects of strong magnetic field. 
In this paper, we investigate the
latter NAE by assuming that the prompt jet of GRB 070110 is driven by neutrino pair annihilation in the NS-disc system. 
Our calculation shows $\eta_{\nu\bar{\nu}}> 1.2 \times 10^{-3}$ under the estimated accretion rate $\dot{M}\simeq 0.04\rm M_{\odot}\cdot s^{-1}$.
Independent of the detailed accretion disc models, our result shows that the magnetic field may play an important role 
in the neutrino annihilation process on the hyperaccreting magnetized accretion disc.
Compared with the theoretical value of $\eta_{\rm \nu \bar{\nu}}$ of the non-magnetized BH-disc system,
the NAE should increase significantly in the case of the NS-disc system if the GRB is powered by magnetar-disc system.
\end{abstract}

\section{Introduction}
Gamma-ray bursts (GRBs) are the brightest events in the universe after the Big Bang.
In general, GRBs can be divided into four stages (Kumar \& Zhang 2015):
(i) a compact binary merger or massive star collapse forms a compact star¨Caccretion disc system (so-called central engine);
(ii) intermittent ultrarelativistic jets are launched from the central engine;
(iii) energy dissipation in these relativistic jets generates gamma-ray emission;
(iv) the interaction between the jets and the surrounding medium results in multiband afterglows.
The central compact star can be a black hole (BH; e.g., Eichler et al. 1989; Narayan, Paczynski, \& Piran
1992; Woosley 1993) or a neutron star (NS; e.g., Usov 1992; Dai \& Lu 1998a,b; Zhang \& M¨¦sz¨¢ros 2001).
The neutrino annihilation process (Popham, Woosley, \& Fryer 1999) and the Blandford-Znajek (BZ) process (Blandford \& Znajek 1977) are the two leading models
used to explain the generation of these relativistic jets.

One of these leading models, the neutrino annihilation process,
has been widely studied theoretically (e.g., Popham, Woosley, \& Fryer 1999; Gu, Liu, \& Lu 2006; Lei et al. 2009)
and using simulations (e.g., Harikae, Kotake, \& Takiwaki 2010;  Zalamea \& Beloborodov (2011); Just et al. 2016; see Liu, Gu, \& Zhang 2017 for review).
According to the the published results, the values of neutrino annihilation efficiency (NAE) under the BH-disk system are mutually corroborated (see Section 4).
However, there is still uncertainty regarding NAE of
NS-accretion disc (NS-disc) system because of the complicated
the microphysics processes and the effects of strong magnetic fields
(Xie, Huang, \& Lei 2007; Xie et al. 2009; Lei et al. 2009; Zhang \& Dai 2009, 2010).
 Until now, there have been no constraints on the
NAE from GRB observations as a result of the unknown launch
mechanism of relativistic jets..

The premise of solving this problem is to determine the type of the central compact star.
In some GRB afterglows, X-ray plateaus can be followed by a very steep decay (e.g. $t^{-9}$, the so-called `internal plateau'; see Figure 1).
This feature indicates that the GRB central engine remains active for some time after the prompt emission is over, and then suddenly shuts down.
Therefore, the internal plateaus are difficult to explain by using the scenario of BH central engines.
It is wildly believed that supramassive strongly magnetized NSs (also called magnetars), which are the central engines of these GRBs, need to be invoked  (Fan \& Xu 2006; Gao \& Fan 2006).
The spin down radiation of the supramassive NS powers the X-ray internal plateau. The transition from the supramassive NS to the BH through the gravitational
collapse after losing rotation energy naturally accounts for the steep decay.

Usually, the magnetar model is incompatible with the BZ mechanism \footnote{Some authors also point out that the NS with a stiff equation
of state (EoS), which has an ergosphere, can power jets via the BZ
mechanism (Ruiz et al. 2012). But since the EoS may be soft  (Margalit \& Metzger 2017),
and plasma is coupled to magnetic field in NSs, we believe that this scenario need to be further studied.}.
Therefore, if it is true that the magnetar is the central engine of a GRB, we may calculate the NAE by using the observation data of this GRB.
In this paper, we find that GRB 070110 is a potential candidate which can be used to address this interesting question.
In Section 2, we present the properties of GRB 070110. We then calculate the NAE $\eta_{\rm \nu \bar{\nu}}$ of this NS-disk system in Section 3.
We compare the NAE of the NS-disc and BH-disc systems in Section 4.
We interpret the results of the comparison in Section 5.
Finally, we give a summary in Section 6. Throughout this paper, a concordance cosmology with
parameters $H_{0} = 70 \rm km s^{-1} Mpc^{-1}$, $\Omega_{\rm M} = 0.30$ and $\Omega_{\Lambda}= 0.70$ is adopted.

\section{The properties of GRB 070110}\label{sec.2}
The \emph{Swift} Burst Alert Telescope (BAT) detected GRB 070110
on 2007 January 10 (Troja et al. 2007). Its duration time $T_{90}$ ($15\;\rm keV, 150\; \rm keV$) is $\sim 89\rm s$
and redshift $z$ is $2.352$. The gamma-ray fluence  ($15\; \rm keV, 150\; \rm keV$) is $\sim 1.8 \times 10^{-6}\;\rm erg\cdot cm^{-2}$.
Then one can obtain the isotropic prompt emission energy $E_{\rm \gamma,iso}\simeq 3.1\times 10^{52}\;\rm erg$ (Du et al. 2016).

In Figure 1, it is clear that a near flat plateau is followed by a steep decay (red line).
As mentioned in the introduction, this internal plateau implies that a hyperaccreting strong magnetized NS central engine is required.
Besides, in the X-ray afterglow of GRB 070110, a bump corresponding to the fall-back BH accretion after the central supramassive NS collapsing to the BH was proposed by Chen et al. (2017).
For these two reasons, we believe that the central object
of GRB 070110 is a supramassive magnetar (hereafter Mag07).

The break time $t_{b}(1+z)$ of the internal plateau is $\sim 2.0 \times 10^{4}\;\rm s$  in observer frame.
After considering the Galactic extinction, the flux of X-ray plateau in $0.3\-10\;\rm keV$  is $F_{\rm x,pla}\sim 1.3 \times 10^{-11}\; \rm erg\;cm^{-2}\;s^{-1}$.
So the isotropic energy of the X-ray plateau in source frame is
\begin{eqnarray}
E_{\rm X,iso,pla}={4\pi D_{\rm L}^{2}}F_{\rm x,pla}t_{b}\simeq 4.3\times 10^{51}\;\rm erg,
\end{eqnarray}
where $D_{\rm L}$ is the the luminosity distance.
Without jet break feature, the jet opening angle only can be constrained as $\theta_{j} > 7.4^{\circ}$ according to the
last observed point $(t_{j}\sim 25 \;\rm d)$ of the X-ray afterglow (Du et al. 2016).
Considering the correction of the jet opening angle, the total energy of prompt emission is
\begin{eqnarray}
E_{\rm \gamma}=E_{\rm \gamma,iso}(1-\cos\theta_{j})>2.5\times 10^{50}\;\rm erg.
\end{eqnarray}
In principle, following the equations and method of Yost et al. (2003), we can obtain the total kinetic energy of GRB jets $E_{\rm K,jet}$.
However, when this method is used to fit the data, some of the parameters are degenerate, we should choose a set of seemingly reasonable parameter values by hand.
Note that $0.01-0.1$ is the typical value of prompt emission efficiency predicted by the matter-dominated jet (Kumar \& Zhang 2015).
Here, we conservatively take the prompt emission efficiency as $0.1$, since the NAE is inversely proportional to the prompt emission efficiency.
So the kinetic energy of the jet is
\begin{eqnarray}
E_{\rm K,jet}=E_{\rm \gamma}/0.1= 2.5\times 10^{51}\;\rm erg,
\end{eqnarray}
and the total jet power iis
\begin{eqnarray}
L_{\rm jet}= E_{\rm K,jet}(1+z)/T_{\rm 90}=9.2\times 10^{49}\; \rm erg\cdot s^{-1}.
\end{eqnarray}

Based on the above results, we discuss how to constrain the properties of Mag07.
Unlike the prompt emission, the energy injection of a magnetar is approximately isotropic.
So the spin-down luminosity $L_{\rm sd}$ can be expressed as
\begin{eqnarray}\label{Eq1}
\eta_{\rm X,pla}L_{\rm sd}=E_{\rm X,iso,pla}/t_{\rm b}\simeq 7.2\times 10^{46}\;\rm erg\cdot s^{-1},
\end{eqnarray}
where $\eta_{\rm X,pla}$ is the X-ray radiation efficiency of the spin-down power.


The collapse of a magnetar into a black hole should occur when a considerable amount of rotation energy is lost.
So the spin-down timescale $\tau$  should be close to the break time $t_{\rm b}$. We assume $\tau=t_{\rm b}$, and we have
\begin{eqnarray}\label{Eq2}
\frac{2\mathcal{\pi}^{2}I}{P^{2}}=L_{\rm sd}t_{\rm b},
\end{eqnarray}
where $I$ and $P$ are the moment of inertia and rotation period of Mag07, respectively.
Theoretically, when the mass of Mag07 $M_{\rm NS}$ equals the maximum mass $M_{\rm max}$ that the star can support, the collapse will occur.
The critical mass $M_{\rm max}$ depends on the equation of state (EoS) and rotational period $P$ of NSs.
According to the observation of the NS binary merger (GW170817/GRB 170817A/kilonova AT2017gfo),
the upper limit on rest mass of NSs is constrained as $M_{\rm res,max}\leq 2.2\;\rm M_{\odot}$ (Margalit \& Metzger 2017)
\footnote{A stiff EoS still has the possibility of existence (Yu, Liu, \& Dai 2018).}.
Here, we adopt the parameters in the EoS APR4 (Read et al. 2009) that $M_{\rm res,max}=2.2\;\rm M_{\odot}$, $I=2.1\times 10^{45}\;\rm g\cdot cm^{2}$ and star radius $R=11.0\;\rm km$.

The spin-down luminosity of Mag07 is
\begin{eqnarray}\label{Eq3}
L_{\rm sd}=\frac{8\mathcal{\pi}^{4}B_{\rm eff}^{2}R^{6}}{3c^{3}P^{4}},
\end{eqnarray}
where $B_{\rm eff}$ is the effective dipole magnetic field strength of Mag07, and $c$ is the speed of light.

In combination with equations (5) and (6), one has
\begin{eqnarray}
P=\left (\frac{\eta_{\rm X,pla}}{7.2\times10^{47}\;\rm erg\cdot s^{-1}} \cdot \frac{2\mathrm{\pi }^{2}I}{t_{\rm b}} \right )^{1/2}.
\end{eqnarray}
Combinating equations (5), (6), and (7) gives
\begin{eqnarray}
B_{\rm eff}=\left (\frac{\eta_{\rm X,pla}}{7.2\times10^{47}\;\rm erg\cdot s^{-1}} \cdot \frac{3c^{3}I^{2}}{2R^{6}t_{\rm b}^{2}} \right )^{1/2}.
\end{eqnarray}
The dependence of period $P$ and magnetic field $B_{\rm eff}$ on the radiation efficiency $\eta_{\rm X,pla}$ is also shown in Figure 2.
As we can see in the upper panel of Figure 2, in addition to having to be less than $1$, since there is a breakup spin period $0.96\;\rm ms$ (dash line) for NSs (Lattimer \&
Prakash 2004), $\eta_{\rm X,pla}$ has a lower limit $0.01$.
In the lower panel of Figure 2, corresponding to $\eta_{\rm X,pla}\in (0.01, 1)$,
the range of $B_{\rm eff}$ is ($6.3\times 10^{14}\;\rm Gs, 6.3\times 10^{15}\;\rm Gs$).


\begin{figure}
\centering
\includegraphics[angle=0,scale=0.28]{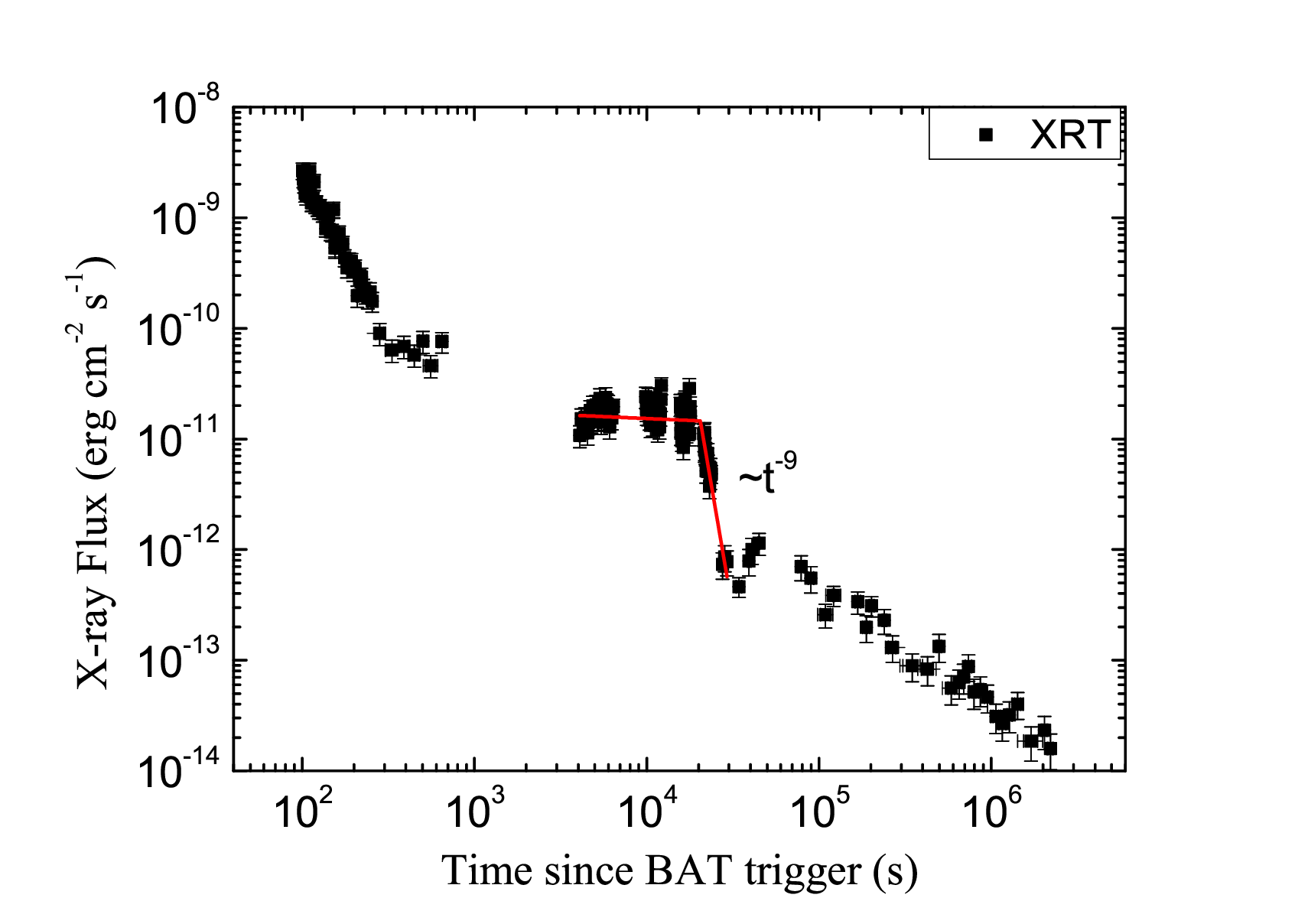}
\caption{The X-ray (black squares) lightcurve of GRB 070110. The solid line is the empirical smooth broken power law function fitting.\label{Fig.1}}
\end{figure}

\begin{figure}
\centering
 \includegraphics[angle=0,scale=0.28]{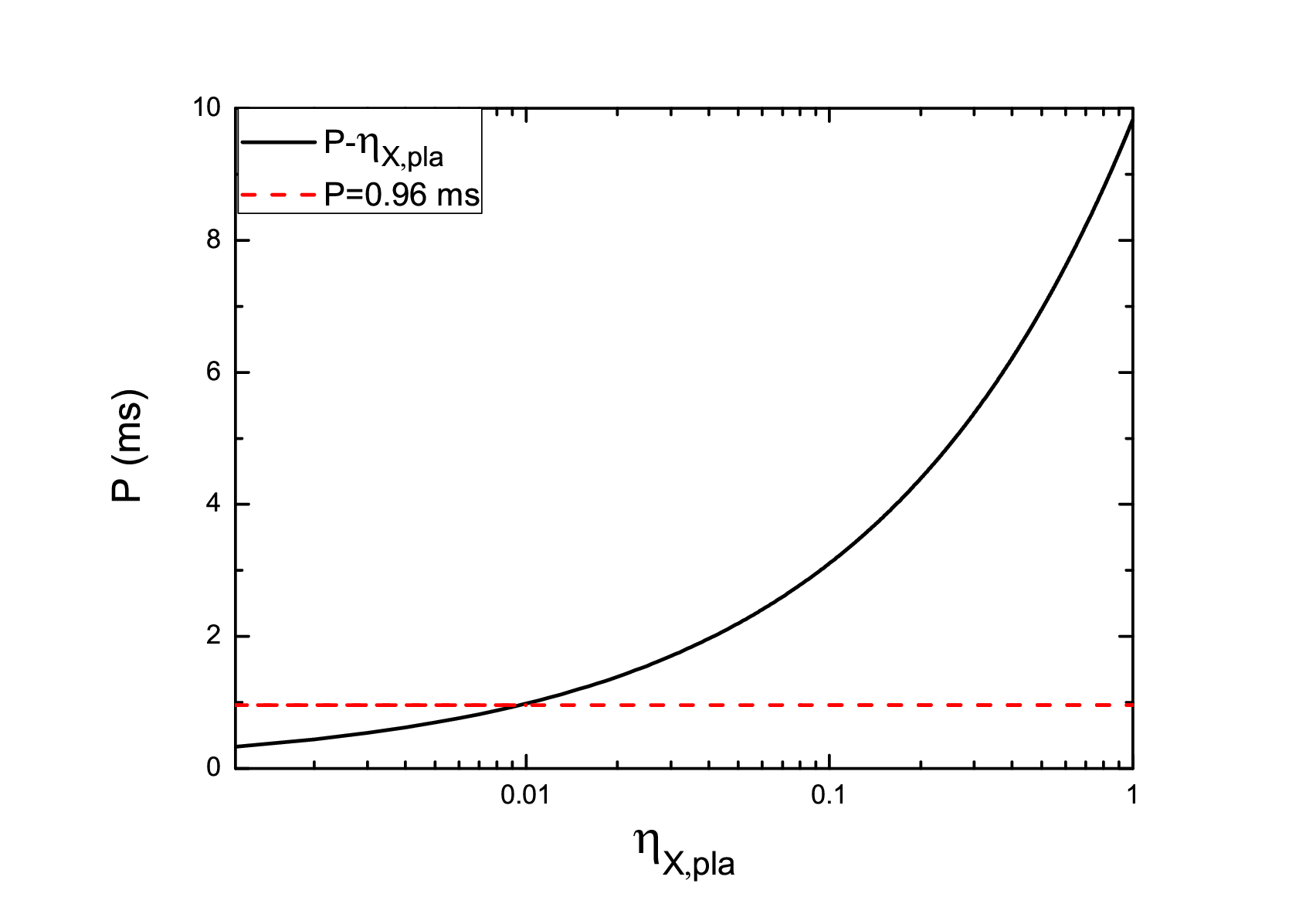}
 \includegraphics[angle=0,scale=0.28]{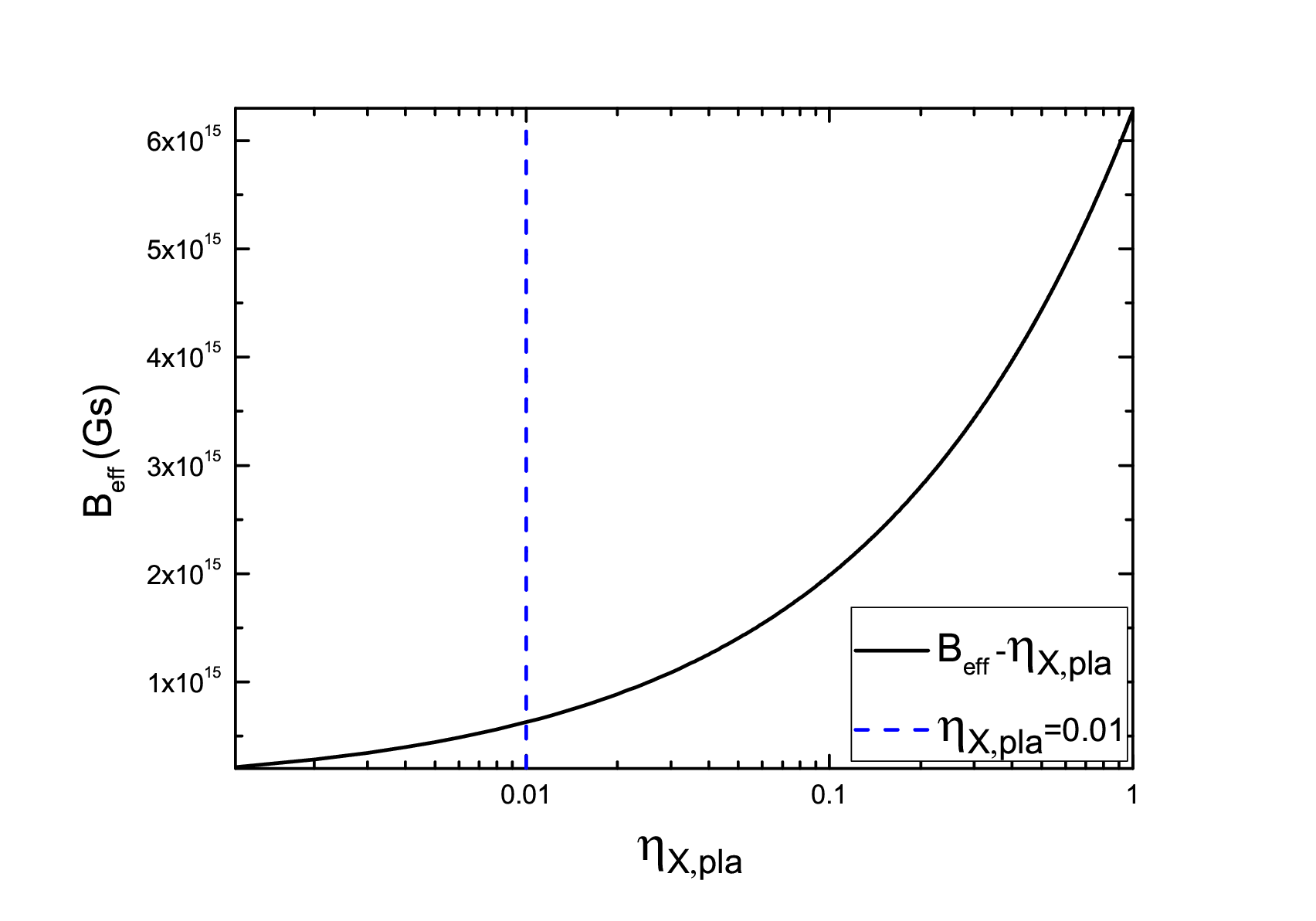}
\caption{The dependence of period $P$ and magnetic field $B_{\rm eff}$ on the radiation efficiency $\eta_{\rm X,pla}$ (solid lines).
In the upper panel, there is a breakup spin period (dash line) for NSs (Lattimer \& Prakash 2004) that the lower limit of the spin period is $0.96\;\rm ms$.
Therefore, the radiation efficiency $\eta_{\rm X,pla}$ has a lower limit $0.01$.
In the lower panel, because $\eta_{\rm X,pla}$ is in $(0.01, 1)$, the ranger of magnetic field $B_{\rm eff}$ is ($6.3\times 10^{15}\;\rm Gs, 6.3\times 10^{15}\;\rm Gs$).\label{Fig.2}}
\end{figure}

\section{Neutrino annihilation efficiency}\label{sec.3}
Observations show that the mass distribution of the NSs in the Milky way is
usually very homogeneously in the range of $1.2-1.6\;\rm M_{\odot}$.
The average mass of these NSs is close to $1.4\;\rm M_{\odot}$ (Zhang et al. 2011).
So we assume the mass of the protomagnetar of GRB 070110 is $M_{\rm pro}=1.4 \;\rm M_{\odot}$.
After accretion, the magnetar mass increases from $1.4\;\rm M_{\odot}$ to $M_{\rm NS}=M_{\rm res,max}+\Delta m$,
where $\Delta m$ is the mass correction after considering the centrifugal force.
The gravitational force of the extra mass should be balanced by the centrifugal force, so we have \footnote{A more accurate method can be referred to Lyford, Baumgarte, \& Shapiro (2003).}
\begin{eqnarray}\label{Eq4}
\frac{G\Delta m}{R^{2}}=\frac{4\mathcal{\pi} R}{P^{2}},
\end{eqnarray}
where $G$ is Newtonian gravitational constant.
For $P>0.96\;\rm ms$, there is $\Delta m<0.4\;\rm M_{\odot}$.
This result is also consistent with the conclusion of Breu \& Rezzolla (2016) that the maximum critical mass of a uniformly rotating NS can increase
at most by $20$ percent of the upper limit on rest mass.
Accordingly, the total mass of accretion disk is $M_{\rm dis}=M_{\rm NS}-M_{\rm pro}<1.2\;\rm M_{\odot}$, and the accretion rate is $\dot{M}=M_{\rm dis}/T_{\rm 90}\simeq 0.04\;\rm M_{\odot}\cdot s^{-1}$.
The neutrino annihilation efficiency $\eta_{\rm \nu \bar{\nu}}$ can be expressed as
\begin{eqnarray}\label{Eq5}
\eta_{\rm \nu \bar{\nu}}=\frac{L_{\rm jet}T_{90}}{(1+z)M_{\rm dis}c^{2}}\geq 1.2\times 10^{-3}.
\end{eqnarray}

\section{The comparison between the NS-disk system and the BH-disk system}\label{sec.4}
To see what $\eta_{\rm \nu \bar{\nu}}\geq 1.2\times 10^{-3}$ means, we compare this value to that of the BH-disk system.
In order to exclude unnecessary interferences, the angular momentum, mass and accretion rate of the BH are the same as that of Mag07.
Since $P>0.96\;\rm ms$, the dimensionless spin of the BH is
\begin{eqnarray}\label{Eq6}
a_{\ast}\simeq\frac{2\pi Ic}{GPM_{\rm res,max}^{2}}<0.3.
\end{eqnarray}

As mentioned in the introduction, the neutrino annihilation luminosity under BH-disk system is wildly discussed.
Some analytic results are shown as follows.

(i) By fitting the results of Popham et al. (1998), Fryer et al. (1999) obtain an approximate formula, i.e.
\begin{eqnarray}\label{Eq7}
\log L_{\rm \nu \bar{\nu}}(\rm erg\cdot s^{-1})&\simeq& 53.4 + 3.4a_{\ast} + 4.89\log \dot{m}\nonumber \\
&<& 46.9,
\end{eqnarray}
where $\dot{m}=\dot{M}/(1\; \rm M_{\odot}\cdot s^{-1})$ is the dimensionless accretion rate.
The second line of equation (11) uses $a_{\ast}<0.03$ and $\dot{m}=0.04$, which is the same below.

(ii) Similarly, by fitting results in Xue et al. (2013), there is (Liu, Gu, \& Zhang 2017)
\begin{eqnarray}\label{Eq8}
\log L_{\rm \nu \bar{\nu}}(\rm erg\cdot s^{-1})&\simeq& 49.5 + 2.45a_{\ast} + 2.17\log \dot{m}\nonumber \\
&<& 46.9.
\end{eqnarray}

(iii) When taking the effect of BH mass into consideration, the annihilation luminosity is (Liu et al. 2016)
\begin{eqnarray}\label{Eq9}
\log L_{\rm \nu \bar{\nu}}(\rm erg\cdot s^{-1})&\simeq& 52.98 + 3.88a_{\ast}-1.55\log m_{\rm NS}\nonumber \\
 + 5.0\log \dot{m} &<& 45.9,
\end{eqnarray}
where $m_{\rm NS}=M_{\rm NS}/\rm M_{\odot}$.

(iv) Lei et al. (2017) also develop a formula to calculate the luminosity of neutrino annihilation, i.e.
\begin{eqnarray}\label{Eq10}
L_{\rm \nu\bar{\nu}} &=&  L_{\rm \nu\bar{\nu},ign} \left[\left(\frac{\dot{m}}{\dot{m}_{\rm ign}}\right)^{-\alpha_{\rm \nu\bar{\nu}}}+
\left(\frac{\dot{m}}{\dot{m}_{\rm ign}}\right)^{-\beta_{\rm \nu\bar{\nu}}}\right]^{-1} \nonumber\\
&&\times \left[1+\left(\frac{\dot{m}}{\dot{m}_{\rm ign}}\right)^{\beta_{\rm \nu\bar{\nu}}-\gamma_{\rm \nu\bar{\nu}} }\right]^{-1},
\end{eqnarray}
where
\begin{eqnarray}\label{Eq11}
\Bigg \{ \begin{array}{ll}
L_{\rm \nu\bar{\nu},ign}=10^{48.0+0.15a_{\ast}}\left (\frac{m_{\rm NS}}{3}\right)^{{\rm log}\left (\dot{m} / \dot{m}_{\rm ign}\right )-3.3} \rm erg~s^{-1}\\
\alpha_{\rm \nu\bar{\nu}}=4.7, \beta_{\rm \nu\bar{\nu}}=2.23, \gamma_{\rm \nu\bar{\nu}}=0.3\\
\dot{m}_{\rm ign}=0.07-0.063a_{\ast}, \dot{m}_{\rm trap}=6.0-4.0a_{\ast}^{3},\\
 \end{array}
\end{eqnarray}
and $\dot {m}_{\rm ign}$ and $\dot {m}_{\rm trap}$ are the dimensionless igniting and trapping accretion rates, respectively.
According to equations (\ref{Eq10}) and (\ref{Eq11}), one can obtain $L_{\rm \nu\bar{\nu}}<2.0\times 10^{47}\;\rm erg\cdot s^{-1}$ under $a_{\ast}<0.03$ and $\dot{m}=0.04$.

All these above equations show a similar result $L_{\rm \nu\bar{\nu}}\; \sim 10^{47}\rm erg\cdot s^{-1}$.
If one believes that this value is right, then it means the NAE on the non-magnetized accretion disc around the BH is at least
two orders of magnitude lower than that of the magnetar case under $\dot{M}\simeq 0.04\; \rm M_{\odot}\cdot s^{-1}$.

\section{The explanation of the difference of the comparison}\label{sec.5}
It is certain that the differences between the NS-disc system and the BH-disc system lead to the different NAEs.
In both systems, mass falls by the way of neutrino dominated accretion flow (Narayan, Paczynski, \& Piran 1992).
The main differences should be: (1) NSs have solid surfaces, but BHs are not;
(2) Mag07 have strong magnetic field, so the accretion disc is magnetized, but situations under the BH-disc systems considered in section 4 are just the opposite.

For the first difference, two extra channels that the annihilation of neutrinos emitted from the NS surface
and annihilation between the neutrinos emitted from the accretion disc and the NS surface
will make contributions to the total NAE. This enhancement depends on the accretion geometry (Zhang \& Dai 2010).
Considering the stability of the accretion disc, magnetic pressure should not be greater than the ram pressure of the accretion flow (also can see Figure 11 of Zhang \& Dai (2010)).
So here, the effect of funnel accretion flow is ignored.
These two extra channels may increase the NAE by one order of magnitude (Zhang \& Dai 2009).
The problem of excessive NAE under the NS-disc system is unsolved.

For the second difference, the energy deposition through $\nu\rightarrow \nu +e^{+}+e^{-}$ and
$\bar{\nu}\rightarrow \bar{\nu} +e^{+}+e^{-}$ in the strong magnetic field will also contribute to the jet power.
But, since the magnetic field of Mag07 is $\sim 10^{15}\;\rm Gs$ (see Figure 2), and the mass accretion rate is $\dot{M}=0.04\;\rm M_{\odot}\cdot s^{-1}$,
this enhancement should be smaller than one order of magnitude (see Figure 7 of Zalamea \& Beloborodov (2011)).
In this sense, the NAE we have calculated previously should be called ``equivalent neutrino annihilation efficiency".
But this effect is still not enough to improve the NAE.

It is necessary to re-examine the energy transfer in accretion process.
In general, to convert gravitational potential energy into thermal energy in an accretion disc, a large viscosity coefficient is needed.
This large viscosity may be induced by the magneto-rotational instability of small-scale magnetic field.
However, the viscosity will lead to the decrease of the surface density of the accretion disc.
That's to say, the magneto-rotational instability will, in turn, inhibit the generation of thermal energy.
Therefore, the thermal luminosity is not sensitive to the viscosity coefficient (Kato, Fukue, \& Mineshige 1998).
When there is a large-scale magnetic field, this situation may be relieved.
According to the flux conservation, the magnetic field on the accretion disc satisfies $B\propto r^{-1}$,
where $r$ is the radius of the disk.
A large-scale magnetic field whose magnetic-pressure gradient toward the outside of the accretion disc may prevent the falling motion of disc matter,
as will as the reduction of the surface density of the disc. So, the thermalization of the accretion disc will become stronger under this situation.
Lei et al. (2009) show a similar result that the strong magnetic field will increase the density of the accretion disc and the NAE from numerical aspect.

Therefore, we get the following conclusion: the large-scale magnetic field in the hyperaccreting accretion disc can effectively enhance the neutrino annihilation luminosity.
This enhancement should be about several tens times larger than that of non-magnetized accretion disc,
such that when the first two enhancements are also taken into consideration, the higher NAE is acceptable.



\section{Summary}\label{sec.6}

The range of mass accretion rate in GRBs is believed to be $0.01\;\rm M_{\odot}\cdot s^{-1} - 10\;\rm M_{\odot}\cdot s^{-1}$.
In this paper, we only calculate the NAE under the accretion rate $\dot{M}\simeq 0.04\;\rm M_{\odot}\cdot s^{-1}$ through a case study: GRB 070110.
In order to get a complete $\eta_{\nu\bar{\nu}}-\dot{m}-B_{\rm eff}$ correlation, more samples like GRB 070110 are required.
On the observation, the internal plateaus are observed both in the afterglows of short and long GRBs. It is possible to achieve this goal in the foreseeable future.
In our case, $\eta_{\nu\bar{\nu}}> 1.2 \times 10^{-3}$ is two order of magnitude larger than that of BH central engine.
The different accretion geometry between the NS-disc system and the BH-disc system may be not enough to explain the high NAE of GRB 070110.
Another effect that the large-scale magnetic field in the accretion disc will obviously improve the NAE can supplement the deficiency of the former.

It is worth reminding that, usually, to power a jet through the BZ mechanism, there must be a strong large-scale magnetic field in the accretion disc.
Therefore, our result indicates that BZ mechanism and neutrino annihilation process are both important under the hyperaccreting BH-disc system.
The structure of the jet from this BH-disc system may be very different from that of the jet only launched by the BZ mechanism
or only produced by the neutrino annihilation process. This structural difference may be very important for explaining some special GRBs, e.g., GRB 170817A (Abbott et al. 2017).

\section{Acknowledgement}
We thank the anonymous referee for his/her useful comments.
We acknowledge the use of the public data from the Swift
data archive, and the UK Swift Science Data Center.
This work is supported by the National Natural Science Foundation of China (Grant No. 11275247, and Grant
No. 11335012) and a 985 grant at Sun Yat-Sen University.
F. K. Peng acknowledges support from the
Doctoral Starting up Foundation of Guizhou Normal University 2017 (GZNUD[2017] 33).


\begin{thebibliography}{}

\bibitem[\protect\citeauthoryear{Abbott et al.}{2017}]{2017ApJ...848L..13A} Abbott B.~P., et al., 2017, ApJ, 848, L13


\bibitem[\protect\citeauthoryear{Blandford \& Znajek}{1977}]{1977MNRAS.179..433B} Blandford R.~D., Znajek R.~L., 1977, MNRAS, 179, 433


\bibitem[\protect\citeauthoryear{Breu \& Rezzolla}{2016}]{2016MNRAS.459..646B} Breu C., Rezzolla L., 2016, MNRAS, 459, 646


\bibitem[\protect\citeauthoryear{Chen et al.}{2017}]{2017ApJ...849..119C} Chen W., Xie W., Lei W.-H., Zou Y.-C., L{\"u} H.-J., Liang E.-W., Gao H., Wang D.-X., 2017, ApJ, 849, 119


\bibitem[\protect\citeauthoryear{Dai \& Lu}{1998a}]{1998A&A...333L..87D} Dai Z.~G., Lu T., 1998a, A\&A, 333, L87


\bibitem[\protect\citeauthoryear{Dai \& Lu}{1998b}]{1998PhRvL..81.4301D} Dai Z.~G., Lu T., 1998b, PhRvL, 81, 4301


\bibitem[\protect\citeauthoryear{Du et al.}{2016}]{2016MNRAS.462.2990D} Du S., L{\"u} H.-J., Zhong S.-Q., Liang E.-W., 2016, MNRAS, 462, 2990


\bibitem[\protect\citeauthoryear{Eichler et al.}{1989}]{1989Natur.340..126E} Eichler D., Livio M., Piran T., Schramm D.~N., 1989, Nature, 340, 126


\bibitem[\protect\citeauthoryear{Fan \& Xu}{2006}]{2006MNRAS.372L..19F} Fan Y.-Z., Xu D., 2006, MNRAS, 372, L19


\bibitem[\protect\citeauthoryear{Fryer et al.}{1999}]{1999ApJ...520..650F} Fryer C.~L., Woosley S.~E., Herant M., Davies M.~B., 1999, ApJ, 520, 650


\bibitem[\protect\citeauthoryear{Gao \& Fan}{2006}]{2006ChJAA...6..513G} Gao W.-H., Fan Y.-Z., 2006, ChJAA, 6, 513


\bibitem[\protect\citeauthoryear{Gu, Liu, \& Lu}{2006}]{2006ApJ...643L..87G} Gu W.-M., Liu T., Lu J.-F., 2006, ApJ, 643, L87


\bibitem[\protect\citeauthoryear{Harikae, Kotake, \& Takiwaki}{2010}]{2010ApJ...713..304H} Harikae S., Kotake K., Takiwaki T., 2010, ApJ, 713, 304


\bibitem[\protect\citeauthoryear{Just et al.}{2016}]{2016ApJ...816L..30J} Just O., Obergaulinger M., Janka H.-T., Bauswein A., Schwarz N., 2016, ApJ, 816, L30


\bibitem[\protect\citeauthoryear{Kato, Fukue, \& Mineshige}{1998}]{1998bhad.conf.....K} Kato S., Fukue J., Mineshige S., 1998, bhad.conf


\bibitem[\protect\citeauthoryear{Kumar \& Zhang}{2015}]{2015PhR...561....1K} Kumar P., Zhang B., 2015, PhR, 561, 1


\bibitem[\protect\citeauthoryear{Lattimer \& Prakash}{2004}]{2004Sci...304..536L} Lattimer J.~M., Prakash M., 2004, Sci, 304, 536


\bibitem[\protect\citeauthoryear{Lei et al.}{2017}]{2017ApJ...849...47L} Lei W.-H., Zhang B., Wu X.-F., Liang E.-W., 2017, ApJ, 849, 47


\bibitem[\protect\citeauthoryear{Lei et al.}{2009}]{2009ApJ...700.1970L} Lei W.~H., Wang D.~X., Zhang L., Gan Z.~M., Zou Y.~C., Xie Y., 2009, ApJ, 700, 1970


\bibitem[\protect\citeauthoryear{Liu, Gu, \& Zhang}{2017}]{2017NewAR..79....1L} Liu T., Gu W.-M., Zhang B., 2017, NewAR, 79, 1


\bibitem[\protect\citeauthoryear{Liu et al.}{2016}]{2016ApJ...821..132L} Liu T., Xue L., Zhao X.-H., Zhang F.-W., Zhang B., 2016, ApJ, 821, 132


\bibitem[\protect\citeauthoryear{Lyford, Baumgarte, \& Shapiro}{2003}]{2003ApJ...583..410L} Lyford N.~D., Baumgarte T.~W., Shapiro S.~L., 2003, ApJ, 583, 410


\bibitem[\protect\citeauthoryear{Margalit \& Metzger}{2017}]{2017ApJ...850L..19M} Margalit B., Metzger B.~D., 2017, ApJ, 850, L19


\bibitem[\protect\citeauthoryear{Narayan, Paczynski, \& Piran}{1992}]{1992ApJ...395L..83N} Narayan R., Paczynski B., Piran T., 1992, ApJ, 395, L83


\bibitem[\protect\citeauthoryear{Popham, Woosley, \& Fryer}{1999}]{1999ApJ...518..356P} Popham R., Woosley S.~E., Fryer C., 1999, ApJ, 518, 356


\bibitem[\protect\citeauthoryear{Read et al.}{2009}]{2009PhRvD..79l4032R} Read J.~S., Lackey B.~D., Owen B.~J., Friedman J.~L., 2009, PhRvD, 79, 124032


\bibitem[\protect\citeauthoryear{Ruiz et al.}{2012}]{2012MNRAS.423.1300R} Ruiz M., Palenzuela C., Galeazzi F., Bona C., 2012, MNRAS, 423, 1300


\bibitem[\protect\citeauthoryear{Troja et al.}{2007}]{2007ApJ...665..599T} Troja E., et al., 2007, ApJ, 665, 599


\bibitem[\protect\citeauthoryear{Usov}{1992}]{1992Natur.357..472U} Usov V.~V., 1992, Nature, 357, 472


\bibitem[\protect\citeauthoryear{Woosley}{1993}]{1993ApJ...405..273W} Woosley S.~E., 1993, ApJ, 405, 273


\bibitem[\protect\citeauthoryear{Xie, Huang, \& Lei}{2007}]{2007ChJAA...7..685X} Xie Y., Huang C.-Y., Lei W.-H., 2007, ChJAA, 7, 685


\bibitem[\protect\citeauthoryear{Xie et al.}{2009}]{2009MNRAS.398..583X} Xie Y., Huang Z.-Y., Jia X.-F., Fan S.-J., Liu F.-F., 2009, MNRAS, 398, 583


\bibitem[\protect\citeauthoryear{Xue et al.}{2013}]{2013ApJS..207...23X} Xue L., Liu T., Gu W.-M., Lu J.-F., 2013, ApJS, 207, 23


\bibitem[\protect\citeauthoryear{Yost et al.}{2003}]{2003ApJ...597..459Y} Yost S.~A., Harrison F.~A., Sari R., Frail D.~A., 2003, ApJ, 597, 459


\bibitem[\protect\citeauthoryear{Yu, Liu, \& Dai}{2018}]{2018ApJ...861..114Y} Yu Y.-W., Liu L.-D., Dai Z.-G., 2018, ApJ, 861, 114


\bibitem[\protect\citeauthoryear{Zalamea \& Beloborodov}{2011}]{2011MNRAS.410.2302Z} Zalamea I., Beloborodov A.~M., 2011, MNRAS, 410, 2302


\bibitem[\protect\citeauthoryear{Zhang \& M{\'e}sz{\'a}ros}{2001}]{2001ApJ...552L..35Z} Zhang B., M{\'e}sz{\'a}ros P., 2001, ApJ, 552, L35


\bibitem[\protect\citeauthoryear{Zhang et al.}{2011}]{2011A&A...527A..83Z} Zhang C.~M., et al., 2011, A\&A, 527, A83


\bibitem[\protect\citeauthoryear{Zhang \& Dai}{2010}]{2010ApJ...718..841Z} Zhang D., Dai Z.~G., 2010, ApJ, 718, 841


\bibitem[\protect\citeauthoryear{Zhang \& Dai}{2009}]{2009ApJ...703..461Z} Zhang D., Dai Z.~G., 2009, ApJ, 703, 461






\end{thebibliography}
\end{document}